\documentclass[doublecol]{epl2}
\usepackage{amssymb}
\usepackage{mathrsfs}
\usepackage{amsmath}
\usepackage{flushend}

\title{Heterogeneous nucleation on complex networks with mobile impurities}
\shorttitle{Heterogeneous nucleation on complex networks} 

\author{Chuansheng Shen\inst{1} \and Hanshuang Chen\inst{2}\and Zhonghuai Hou\inst{3}\thanks{ \email{hzhlj@ustc.edu.cn}}}
\shortauthor{C. Shen \etal}

\institute{
  \inst{1} Department of Physics, Anqing Normal University, Anqing, 246011,
  China \\
  \inst{2} School of Physics and Material Science, Anhui University, Hefei,
230039, China \\
\inst{3} Hefei National Laboratory for Physical Sciences at
Microscales \& Department of Chemical Physics, University of
 Science and Technology of China, Hefei, 230026, China }


\abstract{We study the heterogeneous nucleation of Ising model on
complex networks under a non-equilibrium situation where the
impurities perform degree-biased motion controlled by a parameter
$\alpha$. Through the forward flux sampling and detailed analysis on
the nucleating clusters, we find that the nucleation rate shows a
nonmonotonic dependence on $\alpha$ for small number of impurities,
in which a maximal nucleation rate occurs at $\alpha=0$
corresponding to the degree-uncorrelated random motion. Furthermore,
we demonstrate the distinct features of the nucleating clusters
along the pathway for different preference of impurities motion,
which may be used to understand the resonance-like dependence of
nucleation rate on the motion bias of impurities. Our theoretical
analysis shows that the nonequilibrium diffusion of impurities can
always induce a positive energy flux that can facilitate the
barrier-crossing nucleation process. The nonmonotonic feature of the
average value of the energy flux with $\alpha$ may be the origin of
our simulation results.}

\begin{document}

\maketitle

\section{Introduction}
Nucleation is an activated process which initiates the decay of a
metastable state into a more stable one \cite{Kashchiev2000} driven
by fluctuation. Many important dynamical processes on real-world
scenarios, such as crystallization\cite{JCP97003634,AMI09001203},
fractures\cite{NAT91000039,PRL97003202}, glass
formation\cite{PRE98005707}, and protein folding\cite{PNAS9510869},
to list just a few, are concerned with nucleation. For many decades,
our understanding of nucleation has been dominated by the classical
nucleation theory (CNT), and it has been applied not only to the
liquid-gas and liquid-solid systems, but also to regular lattices in
Euclidean space. For instance, in two-dimensional lattices, Allen
\emph{et al} discovered that shear can enhance the nucleation rate
and the rate peaks at an intermediate shear rate \cite{JCP08134704}.
Sear found that a single impurity may considerably enhance the
nucleation rate \cite{JPC06004985}. Page and Sear reported that the
existence of a pore may lead to two-stage nucleation, and the
overall nucleation rate can reach a maximum level at an intermediate
pore size \cite{PRL06065701}. In three-dimensional lattices, the
nucleation pathway of the Ising model has also been studied by Sear
and Pan \cite{JCP08164510,JPC0419681}. In addition, the validity of
CNT has been tested in other Euclidean space
\cite{EPJ98000571,JCP99006932,JCP00001976,PRE05031601,PRL09225703,PRE10030601,PRE10011603}.

 Since many real systems can be properly modeled by network-organized
structure \cite{RMP02000047,AIP02001079,SIR03000167}, it is thus an
interesting topic to explore nucleation process in complex networks.
 Recently, we have studied nucleation
dynamics of the Ising model on scale-free (SF) networks
\cite{PRE11031110}, Erd\"{o}s-R\'{e}nyi (ER)
networks\cite{Chaos13013112} and modular networks
\cite{PRE11046124}. We found that, for homogeneous nucleation on SF
networks, many small isolated nucleating clusters emerge at the
early stage of the nucleation process, until suddenly they form the
critical nucleus through a sharp merging process, and the nucleation
rate decays exponentially with network size. For homogeneous
nucleation on ER networks, there always exists a dominant nucleating
cluster to which relatively small clusters are attached gradually to
form the critical nucleus. For modular networks, as the network
modularity worsens the nucleation undergoes a transition from a
two-step to one-step process and the nucleation rate shows a
nonmonotonic dependence on the modularity. For heterogeneous
nucleation, target impurities are shown to be much more efficient to
enhance the nucleation rate than random ones. However, in our
previous work, impurities are considered to be fixed in some nodes
with the new phase. As we know, mobility is a ubiquitous feature of
real systems \cite{NAP2007276,PRL07148701, PhysRevLett.100.044102,
PhysRevE.83.025101, PhysRevE.87.032814}, including the mobility of
impurities, and may drastically influence the dynamical evolution.
For example, it has been reported that mobility promotes
synchronization \cite{PNAS104979,PhysRevLett.110.114101}, enhances
signal response \cite{EPL1338004}, affects contagion
processes\cite{NAP2011581}, and tunes biodiversity
\cite{NAT20071046}. In addition, impurities may be caused by the
vacancy defects with no interaction, not with new phase. How the
impurity motion would influence the nucleation rate and pathway is
still an open question. Motivated by this, we will study the
different roles of the motion in the formation of nucleating
clusters, which can reveal the nucleation pathways of the Ising
model in the underlying networks.

In the present work, we adopt the recently proposed forward flux
sampling (FFS) \cite{PRL05018104} approach, which is efficient and
easy to implement to study rare events, and employ SF networked
Ising model. Ising model is a paradigm for many phenomena in
statistical physics and widely used to study the nucleation process.
By introducing degree-biased random walks for impurities on the
network, we find that for small number of impurities the nucleation
rate shows a nonmonotonic dependence on the bias parameter of the
motion of impurities, in which a maximal nucleation rate occurs at
the situation where the impurities perform random motions.
Furthermore, we show that there are different properties of the
nucleating clusters along the pathway corresponding to different
impurities bias motions.

\section{Model and method} \label{sec2}

\subsection{ Model}

We consider the non-equilibrium Ising model with mobile impurities
on complex networks consisting of $N$ normal nodes and $w$ defect
nodes called impurities. Each normal node is endowed with a spin
variable $s_i$ that can be $+1$ (up), or $-1$ (down), and each
defect node is endowed with spin $0$ (impurity). The Hamiltonian of
the system is given by
\begin{equation}
H =  - J\sum\limits_{i < j} {A_{ij} s_i s_j } - h\sum\limits_i{s_i},
\label{eq1}
\end{equation}
where $J$ is the coupling constant and $h$ is the external magnetic
field. For convenience, we set $J=1$ in the following discussions.
The elements of the adjacency matrix of the network take $A_{ij} =
1$ if nodes $i$ and $j$ are connected and $A_{ij} =0$ otherwise. The
degree, that is the number of neighboring nodes, of node $i$ is
defined as $ k_i = \sum\nolimits_{j = 1}^N {A_{ij} }.$  Notice that,
without defect nodes, there exist a number of simulations and
analytical results for the Ising model in ER and SF networks
\cite{PRE02016104,PRE04067109,PRE06036108,PRE09051127,PRE10011102}.

The dynamical evolution of our model has two ingredients: Spin-flip
and Impurity diffusion. In each time step, we attempt to perform the
following two types of dynamics. 1) Spin-flip: we randomly chose a
normal node and attempt to flip its spin according to the Metropolis
acceptance probability $\min (1,e^{ - \beta \Delta E} )$
\cite{Lan2000}, where $\beta = 1/(k_B T )$ with the Boltzmann
constant $k_B$ and the temperature $T$, and $\Delta E$ is the energy
change due to the flipping process; 2) Impurity diffusion: After a
spin of a normal node has been attempted to be flipped, we then
randomly choose a defect node(impurity node) $i$ and exchange the
spin $s_i$ with that $s_j$ of one of the nearest neighboring normal
nodes $j$ according to the probability
\begin{eqnarray}\label{eq2}
{p_{i \leftrightarrow  j}} & = & D\frac{{k_j^\alpha
}}{{\sum\nolimits_{l \in \mathscr{N}(i), {s_l} \neq 0} {k_l^\alpha }
}}
\end{eqnarray}
Here $D$ is the diffusion constant, the sum is taken over all the
nearest neighboring normal nodes of $i$, and $\alpha$ is a tunable
parameter which biases the impurities' motion either towards
low-degree nodes ($\alpha<0$) or towards hubs ($\alpha>0$). For
$\alpha=0$, we recover the standard (unbiased) random walk.

In general, with the increment of $T$ the system undergoes a
second-order phase transition from an ordered state to a disordered
one at the critical temperature $T_c$. Below $T_c$ the system
prefers to be in a state with all spins up or down. Given a small
external field, one of these two states will become metastable, and
if initiated predominantly in this metastable state, the system will
remain for a significantly long time before it undergoes a
nucleation transition to the thermodynamically stable state. We are
interested in the rate and pathways for this transition.

\subsection{FFS method}

The FFS method has been successfully used to calculate rate
constants and transition paths for rare events in equilibrium and
non-equilibrium systems
\cite{JCP08134704,JPC06004985,PRL06065701,PRL05018104,JCP07114109,JCP06024102}.
 This method uses a series of interfaces in phase space
between the initial and final states to force the system from the
initial state $A$ to the final state $B$ in a ratchetlike manner.
First, we define an order parameter $\lambda(x)$, where $x$
represents the phase-space coordinates, such that the system is in
state $A$ if $\lambda(x)<\lambda_0$ and state $B$ if $\lambda(x)>
\lambda_M$, while a series of nonintersecting interfaces
$\lambda_i$($0<i<M$) lie between states $A$ and $B$, such that any
path from $A$ to $B$ must cross each interface without reaching
$\lambda_{i+1}$ before $\lambda_i$. The transition rate $R$ from $A$
to $B$ is calculated as
\begin{eqnarray}\label{eq3}
R = \bar \Phi _{A,0} P\left( {\lambda _M |\lambda _0 } \right) =
\bar \Phi _{A,0} \prod\nolimits_{i = 0}^{M - 1} {P\left( {\lambda
_{i + 1} |\lambda _i } \right)}
\end{eqnarray}
where $\bar \Phi _{A,0} $ is the average flux of trajectories
crossing $\lambda_0$ in the direction to $B$. $ P\left( {\lambda _M
|\lambda _0 } \right) =  \prod\nolimits_{i = 0}^{M - 1} {P\left(
{\lambda _{i + 1} |\lambda _i } \right)}$ is the probability that a
trajectory crossing $\lambda_0$ in the direction to $B$ will
eventually reach $B$ before returning to $A$, and $P\left( {\lambda
_{i + 1} |\lambda _i } \right)$ is the probability that a trajectory
which reaches $\lambda_i$, having come from $A$, will reach
$\lambda_{i+1}$ before returning to $A$. For more detailed
descriptions of FFS method, please see Ref. \cite{JPH09463102}.

In this work, we perform Monte Carlo simulation and use FFS to study
nucleation rate and pathways of the non-equilibrium phase from the
metastable spin phase. Specifically, we set $T<T_c$, $h=0.5$ and
start from an initial state with $s=-1$ for most of the spins. We
define the order parameter $\lambda$ as the total number of up spins
in the network. The spacing between adjacent interfaces is fixed at
$3$ up spins. We perform $1000$ trials per interface for each FFS
sampling, from which at least $200$ configurations are saved in
order to investigate the statistical properties along the nucleation
pathway. The results are obtained by averaging over $20$ independent
FFS samplings and $50$ different network realizations.

\section{Results and Discussion}  \label{sec3}

\subsection{ Nucleation rate}

In what follows, we employ a Barab\'{a}si-Albert (BA) SF network,
whose degree distribution follows a power law $p(k) \sim
k^{-\gamma}$ with the scaling exponent $\gamma = 3$
\cite{SCI99000509}.

\begin{figure}
\centerline{\includegraphics*[width=0.95\columnwidth]{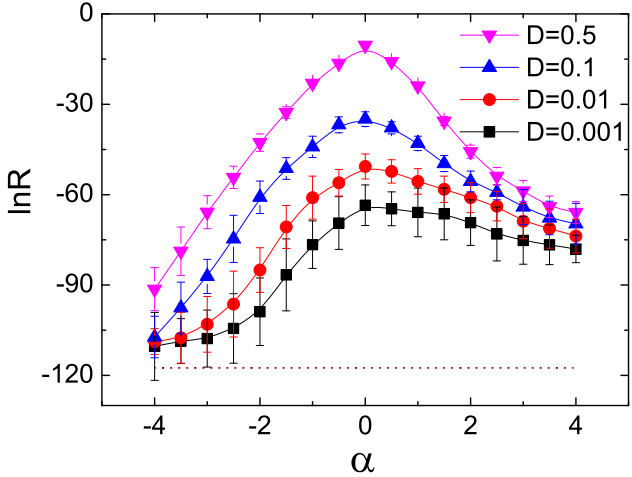}}
\caption{(Color online) The nucleation rate $\ln R$ as a function of
$\alpha$ for different values of the diffusion rate $D$. The dotted
line indicates the result without impurities, i.e., $w=0$.
Parameters are $N = 1000$, the average network degree $\langle k
\rangle =6$, $w=2$, $h=0.7$, $T=2.59$, $\lambda_0 = 130$ and
$\lambda_M = 880$. \label{fig1}}
\end{figure}

To begin, we fix the small number $w=2$ of impurities and vary the
diffusion constant $D$ to investigate how the nucleation rate $R$
(in unit of $MC step^{-1}spin^{-1}$) evolves with controlling
parameter $\alpha$. Figure \ref{fig1} shows the dependence of the
logarithm of the nucleation rate $\ln R$ on $\alpha$ for different
values of $D$. One can observe that $\ln R$ exhibits a
resonance-like behavior with the increment of $\alpha$. That is,
there exists an optimal value of $\alpha$ at $\alpha=\alpha_{opt}$,
corresponding to the maximum $R$. Interestingly, this phenomenon is
robust against the diffusion constant $D$. This result indicates
that random motions of impurities corresponding to $\alpha_{opt}=0$
is more favorable to nucleation than degree-biased motions. In
addition, for any given values of $\alpha$ we find that $\ln R$
increases monotonously with $D$, indicating that impurities mobility
is always in favor of nucleation. That is, the larger the mobility
rate $D$ is, the larger $\ln R$ becomes. The dotted line indicates
the result without impurities, i.e., $w=0$. Obviously, for any given
value of $D$ impurities may considerably enhance the nucleation
rate, which is consistent with \cite{JPC06004985}.

\begin{figure}[h]
\centerline{\includegraphics*[width=0.95\columnwidth]{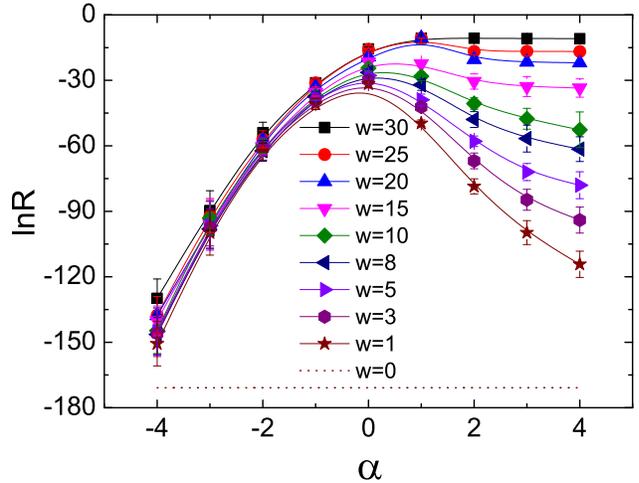}}
\caption{(Color online) $\ln R$ as a function of $\alpha $ for
different values of $w$. The dotted line indicates the result
without impurities, i.e., $w=0$. Other parameters are same as in
Fig.1 except for $D=0.5$ and $h=0.5$. \label{fig2}}
\end{figure}

It is worthy noting that this nontrivial dependence is unobservable
if the number of impurities $w$ becomes relatively large. Figure
\ref{fig2} shows the dependence of $\ln R$ on $\alpha $ for
different values of $w$. Clearly, for small number of impurities,
say $w=1,3,5,8,10$, one can always observe an interesting mobility
induced resonance-like behavior in accordance with Fig.1. While for
big $w$, say $w=15,20,25,30$, $\ln R$ increases monotonously with
$\alpha$ for $\alpha \leq 1$ and then approaches a constant value
for $\alpha>1$. Other values of $w$ have also been investigated; the
qualitative results are the same and not shown here. But for $w=0$
indicated by the dotted line, i.e.,without impurities, $\ln R$ is
considerably less than that of impurities.

\subsection{ Nucleation pathway}
To elucidate the detailed characteristics along the nucleation
pathway for different bias motions, we save lots of configurations
generated by FFS and perform detailed analysis
 on the nucleating clusters, including the relative size of the largest and the second largest cluster, average
degree of the cluster nodes and the number of clusters. According to
CNT, there exists a critical nucleus size $\lambda_c$ of the new
phase, above which the system grows rapidly to the new phase.
Herein, we mainly focus on the nucleation stage where
$\lambda<\lambda_c$. In our simulation, we determine $\lambda_c$ by
computation of the committor probability $P_B$, which is the
probability of reaching the thermodynamic stable state before
returning to the metastable state. As commonly reported in the
literature \cite{JPC0419681,PRE10030601}, the critical nucleus
appears at $P_B(\lambda_c)=0.5$. Since $\lambda_c$ are different for
different bias parameters, we thus introduce $\lambda/\lambda_c$ as
the control parameter.

\begin{figure}
\centerline{\includegraphics*[width=1.0\columnwidth]{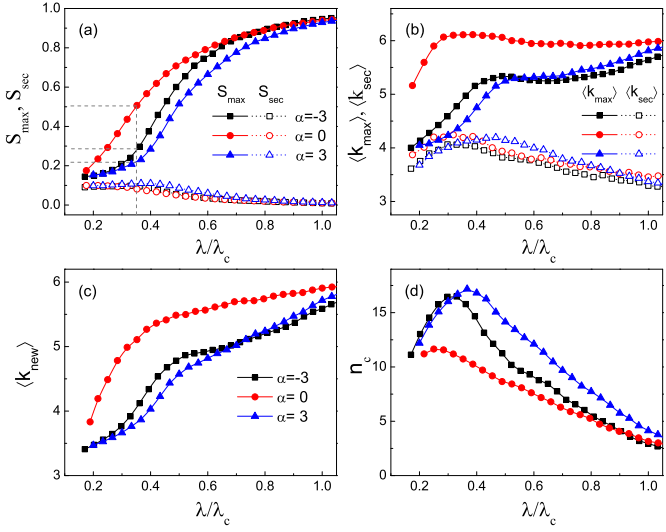}}
\caption{(Color online) (a) The relative size $ S_{max} $, $S_{sec}$
of the largest and the second largest nucleating cluster
respectively, as a function of $\lambda /\lambda _c $. (b) The
average degrees $k_{max}$, $k_{sec}$ of the nodes within the largest
and the second largest nucleating cluster respectively, as a
function of $\lambda /\lambda _c $. (c) and (d) correspond to the
average degree $\langle k_{new}\rangle $ of the nodes inside
nucleating clusters and the number $n_c$ of nucleating clusters
respectively, as a function of $\lambda /\lambda _c$ for $\alpha=-3,
0, 3$, corresponding to motion preferring to low-degree nodes, to
random nodes and high-degree nodes respectively. Symbols for
different motion bias in (b), (c) and (d) are same as in (a). Other
parameters are same as in Fig.1 except for $D=0.5$ and $h=0.5$.
\label{fig3}}
\end{figure}

Following the previous study \cite{Chaos13013112}, we introduce the
relative size $S_{max}$, $S_{sec}$ of the largest and the second
largest nucleating cluster, which are defined as the ratios of the
number of up spins within the largest and the second largest cluster
to the total number of up spins respectively. $S_{max}$ and
$S_{sec}$ (averaged over the ensemble at each interface) as a
function of $\lambda /\lambda _c $ are plotted in figure \ref{fig3}
(a). Clearly, one can see that $S_{max}$  for $\alpha=0$ (solid red
circles) is always larger than those for $\alpha=-3$ (solid black
squares) and $\alpha=3$ (solid blue triangles). Specifically, at
$\lambda /\lambda _c=0.35 $, $S_{max}$ grows already more than $50\%
$ for $\alpha=0$, while it is less than $30\% $ for $\alpha=-3$ and
about $20\% $ for $\alpha=3$, as shown by the dashed gray lines in
figure \ref{fig3}(a). But when $\lambda /\lambda _c=1$ they almost
tend to $100\%$ together. This difference means that for unbiased
random motion $S_{max}$ grows fast at the very beginning following
by a relatively slow increasing, while for biased motion, $S_{max}$
increases slowly at first and then rapidly when approaching the
critical nucleus. From figure \ref{fig3} (a) one can also observe
that the relative size $S_{sec}$ (denoted by the empty symbols) is
greatly less than $S_{max}$, indicating that the nucleation is
dominated by the largest nucleating cluster.

We also plot $\langle k_{max}\rangle $ and $\langle k_{sec}\rangle
$, defined as the average degrees of the nodes within the largest
and the second largest nucleating cluster respectively, as a
function of $\lambda /\lambda _c $ in figure \ref{fig3} (b).
Clearly, $\langle k_{max}\rangle $ for $\alpha=0$ indicated by the
solid red circles, is always larger than those for $\alpha=-3$ and
$\alpha=3$ indicated by the solid black squares and solid blue
triangles respectively. Strikingly, at the early nucleating stage,
$\langle k_{max}\rangle $ grows sharply for the former, while it
grows gradually for the latter. Furthermore, it is found that
$\langle k_{sec}\rangle $ is greatly less than $\langle
k_{max}\rangle $, which suggests again that the largest nucleating
cluster dominates the nucleation.

In addition, we also investigate the average degree $\langle
k_{new}\rangle $ of the nodes inside the new phase and the number
$n_c$ of nucleating clusters, and plot $\langle k_{new}\rangle $ and
$n_c$ as a function of $\lambda /\lambda_c$ in figure \ref{fig3} (c)
and (d) respectively. As shown, $\langle k_{new}\rangle $ increases
monotonically with $\lambda /\lambda _c $ for different $\alpha=-3,
0, 3$, which means the new phase tends to grow from those nodes with
smaller degrees. Nevertheless, for different preference of
impurities motion it shows the distinct features along the
nucleation pathway. For $\alpha=0$, $\langle k_{new}\rangle $ grows
fast at the very beginning following by a relatively slow
increasing. For $\alpha=-3$ and $3$, $\langle k_{new}\rangle $
increases slowly at first and then grows fast until approaching the
critical nucleus. Such a scenario is consistent with figures
\ref{fig3} (a) and (b). From figure \ref{fig3} (d) one can observe
that $n_c$ shows non-monotonically dependence on $\lambda /\lambda
_c $ for different $\alpha$. Especially, the number of clusters for
$\alpha= 0$ is always less than that for $\alpha=-3, 3$. On the
other hand, $n_c$ approaches the same magnitude near the formation
of critical nucleus for three different $\alpha$, which suggests
that it is easier for the critical nucleus comes into being for
unbiased random motion than that for others. This result is also
consistent with the picture shown in figures \ref{fig3} (a) to (c).

\begin{figure}
\centerline{\includegraphics*[width=0.5\columnwidth]{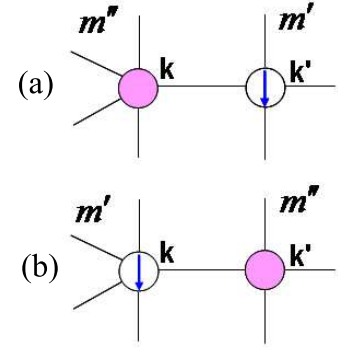}}
\caption{(Color online) Schematic illustration of an impurity
diffusion. (a) indicates an impurity node occupied a $k$-degree node
(filling node) will exchange its present position with a spin on a
$k'$-degree node (arrow node), and (b) denotes the reverse process.
$m'$ and $m''$ denote the average magnetization of a randomly chosen
neighboring normal node and defect(impurity) node respectively.
About their definitions, please see Eqs.(\ref{eq4},\ref{eq5}).
\label{fig4}}
\end{figure}

To further understand the nontrivial effect of the nonequilibrium
diffusion on nucleation, we will evaluate the average energy change
due to the diffusion process between an impurity and a spin. To
begin, let us consider such a process on a link connecting a
$k$-degree node and a $k'$-degree node. As shown in
Fig.\ref{fig4}(a), an impurity occupied $k$-degree node will
exchange its present position with a spin of $k'$-degree node. The
reverse process is shown in Fig.\ref{fig4}(b). Let $q_k$ and $m_k$
denote the probability of a $k$-degree node occupied by an impurity
and the average magnetization of a $k$-degree node occupied by a
spin, respectively. For a spin node on degree uncorrelated networks,
the average magnetization $m'$ of a randomly chosen neighboring node
is written as,
\begin{eqnarray}\label{eq4}
m' = \sum\nolimits_k {\frac{{kP(k)}}{{\left\langle k \right\rangle
}}} {m_k}(1 - {q_k})
\end{eqnarray}
While for an impurity node, the average magnetization of a randomly
chosen neighboring node is written as
\begin{eqnarray}\label{eq5}
m'' = \sum\nolimits_k {\frac{{kP(k)}}{{\left\langle k \right\rangle
}}} {m_{k-1}}(1 - {q_k})
\end{eqnarray}
where the subscript $k-1$ other than $k$ is based on the
consideration that there is no any interaction between an impurity
node and its neighbors. For the case of Fig.\ref{fig4}(a), the
energy change is written as
\begin{eqnarray}\label{eq6}
\Delta E_1  = -J m_{k'} [(k - 1)m'' - (k' - 1)m']
\end{eqnarray}
Analogously, for the case of Fig.\ref{fig4}(b)
\begin{eqnarray}\label{eq7}
\Delta E_2   = -J m_k[(k' - 1)m'' - (k - 1)m']
\end{eqnarray}
The energy change due to a diffusing exchange taking placing on a
$k-k'$ link can be expressed as
\begin{eqnarray}\label{eq8}
\Delta E = \frac{1}{2}\left( {{q_k}{p_{k \to k'}}\Delta {E_1} +
{q_{k'}}{p_{k' \to k}}\Delta {E_2}} \right)
\end{eqnarray}
where ${p_{k \to k'}}$ is the diffusion rate of an impurity from
$k$-degree node to $k'$-degree node. According to our model, it can
be expressed as
\begin{eqnarray}\label{eq9}
p_{k \to k'}  = D \frac{{k'^\alpha  }}{{\sum\nolimits_{k'}
{\frac{{k'P(k')}}{{\langle k\rangle }}} k'^\alpha  }}  = D
\frac{{\langle k\rangle k'^\alpha  }}{{\langle k^{\alpha + 1}
\rangle }}
\end{eqnarray}
Assuming the diffusion is quasi-static process that satisfies the
detailed balance conditions,
\begin{eqnarray}\label{eq10}
{q_k}{p_{k \to k'}} = {q_{k'}}{p_{k' \to k}}
\end{eqnarray}
The requirement can lead to the expression of $q_k$,
\begin{eqnarray}\label{eq11}
 q_k
= \frac{{wP(k)k^{\alpha  + 1} }}{{\sum\nolimits_k {P(k)k^{\alpha  +
1} } }}
\end{eqnarray}
Averaging over all possible links on networks, one arrives at the
average energy change due to a nonequilibrium diffusion process,
\begin{eqnarray}\label{eq12}
\left\langle {\Delta E} \right\rangle  = \sum\limits_{k,k'}
{{l_{kk'}}} \Delta E
\end{eqnarray}
where ${l_{kk'}} = {{{kk'P(k)P(k')} \mathord{\left/
 {\vphantom {{kk'P(k)P(k')} {\left\langle k \right\rangle }}} \right.
 \kern-\nulldelimiterspace} {\left\langle k \right\rangle }}^2}$ is
the probability that a randomly chosen link to connect a pair of
nodes with degree $k$ and $k'$.

Next we will calculate the average magnetization using heterogeneous
mean-field theory. Following Ref.\cite{PLA02000166}, one has
\begin{eqnarray}\label{eq13}
  m_k  = \tanh [\beta Jkm' + \beta h]
\end{eqnarray}
Substituting Eq.(\ref{eq13}) with Eqs.(\ref{eq4}-\ref{eq5}), we
arrive at the self-consistent formulations of $m'$ and $m''$ that
can be numerically calculated.

\begin{figure}
\centerline{\includegraphics*[width=0.75\columnwidth]{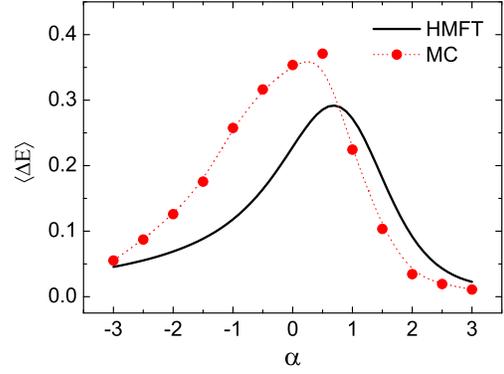}}
\caption{(Color online) The average energy change $\left\langle
{\Delta E} \right\rangle$ as a function of $\alpha$. The solid line
indicates the results of the theoretical prediction, and solid
symbols that of MC. Parameters are same as in Fig.1 except for
$D=1.0$ and $h=0.5$. \label{fig5}}
\end{figure}

Figure \ref{fig5} shows the results of $\left\langle {\Delta E}
\right\rangle$ as a function of $\alpha$, where the solid line
denotes the result obtained from Eq.(\ref{eq12}), and the symbols
that of MC simulations. Clearly, the theory can reproduce
qualitatively well the main characteristic: there exists an optimal
motion bias where the average energy change reaches the maximum.
Furthermore, it is found that $\left\langle {\Delta E}
\right\rangle$ are always larger than zero for any motion bias
$\alpha$, which indicates that the impurity's mobility can always
facilitate the barrier-crossing nucleation process, akin to the drag
effect of nonequilibrium thermodynamic forces conjugated to the
exchange of impurities and spins.

\section{Conclusions} \label{sec4}

In summary, we have studied the heterogeneous nucleation of a
non-equilibrium Ising model with mobile impurities on complex
networks. By introduced a tunable parameter $\alpha$, the impurities
can perform three different bias motions: $\alpha>0$ means that the
impurities prefer to visit the high-degree nodes, $\alpha<0$ the
low-degree nodes, and $\alpha=0$ recovers to completely random
motion. Interestingly, it is found that the nucleation rate is not a
monotonic function of $\alpha$ for small number of impurities, i.e.,
there exists an optimal value of $\alpha=0$, leading to the fastest
nucleation rate. Especially, the optimal value of the controlled
parameter does not change with the variation of the diffusion rate.
To qualitatively understand the underlying mechanism of such a
phenomenon, we have performed heterogeneous mean-field analysis.
Furthermore, we have used the FFS method and analyzed the nucleating
clusters, and found that for different preferences of impurities
motion, the nucleating clusters show the distinct features along the
pathways. On the one hand, for random motion, the largest nucleating
cluster dominates the nucleation, and the average degree of the
nodes inside nucleating clusters grows rapidly at first, while for
motion to the high-degree nodes or to the low-degree nodes, they
grow slowly at the very beginning following by a relatively fast
increasing. On the other hand, the number of nucleating clusters
 for the former is less than that of the latter, especially they decreases to the same magnitude
 at the formation of the critical nucleus. These distinct features may mean different
microscopic mechanisms driving the system towards nucleation. Since
heterogeneous nucleation is essential for many dynamical processes
on real-world scenarios, and mobility is a ubiquitous feature of
real systems, our study may provide a valuable understanding for
many non-equilibrium phase transitions taking place in networked
systems and for effective controlling strategy to the rate of such
processes.

\acknowledgments This work was supported by the National Basic
Research Program of China (2013CB834606) and by the National Natural
Science Foundation of China (Grants No. 21125313, 11205002, 21473165
and No. 11475003). C.S.S. was also supported by Anhui Provincial
Natural Science Foundation (Grant No.1408085MA09).

\bibliographystyle{eplbib}

\end{document}